\documentclass[twocolumn,aps,floats,superscriptaddress]{revtex4}
\usepackage{graphicx,epsfig}
\usepackage{times}
\usepackage{graphics,dcolumn,bm,float}
\usepackage{amssymb,amsmath,rotate,color}
\usepackage[normalem]{ulem}
\usepackage{soul}

\begin{document}

\title{Thermoelectric voltage switching in gold atomic wire junctions}
\author{Alireza Saffarzadeh}
\altaffiliation{Author to whom correspondence should be addressed.
Electronic mail: asaffarz@sfu.ca} \affiliation{Department of
Physics, Payame Noor University, P.O. Box 19395-3697 Tehran, Iran}
\affiliation{Department of Physics, Simon Fraser University,
Burnaby, British Columbia, Canada V5A 1S6}
\author{Firuz Demir}
\affiliation{Department of Physics, Simon Fraser University,
Burnaby, British Columbia, Canada V5A 1S6}
\author{George Kirczenow}
\affiliation{Department of Physics, Simon Fraser University,
Burnaby, British Columbia, Canada V5A 1S6}

\date{\today}

\begin{abstract}
We explore the thermoelectric properties of gold atomic chains
bridging gold electrodes by means of \textit{ab initio} and
semi-empirical calculations, and heuristic reasoning. We predict
the thermoelectric voltage induced by a temperature difference
across such junctions to oscillate, repeatedly changing sign, as a
function of the number of atoms $N$ making up the atomic chain. We
also predict the amplitude of the oscillations to be proportional
to $N$ for long atomic chains. Further we predict the
thermoelectric voltage to change sign, in some cases, if the
junction is stretched without changing $N$. Our predictions apply
regardless of whether regular or irregular electrode grain
boundaries are present and whether the electrodes are symmetric or
asymmetric. Our findings may pave the way to the realization of
new mechanically controllable voltage switches enabling direct
conversion of heat into electricity in energy harvesting
applications.

\end{abstract}
\maketitle

\section{Introduction}
In the past two decades, the formation, evolution, and breaking of
gold atomic chains that bridge a pair of electrodes have been the
subject of many experimental and theoretical studies, motivated by
their fundamental scientific interest and potential applications
as the narrowest possible conducting wires for nanoscale
electronic devices
\cite{Agrait,Smit2003,Vardimon2014,Emberly,Vega,Silva2004,Dreher,Tavazza,Tavazza13,Zheng2015}.
The electric current through such atomic chains can be measured in
scanning tunneling microscopy experiments \cite{Kawahito,Yazdani}.
Also, mechanically controlled break junctions can be used to form
a gold atomic chain between two gold clusters. In this way, atomic
chains up to 2 nm long have been fabricated in ultrahigh vacuum
\cite{Agrait,Ohnishi,Yanson}. The production of very thin
suspended gold nanowires, as thin as a linear atomic chain with
4-10 atoms has also been reported
\cite{Rodrigues1,Rodrigues2,Thijssen,Kizuka}. Moreover, molecular
dynamics simulations of stretched gold nanowires have shown that
the number of atoms in monatomic chains is a function of
temperature and that there is an optimal temperature for creation
of long atomic chains \cite{Cortes-Huerto}.

The length of these atomic chains can be estimated experimentally
from the length of the last conductance plateau
\cite{Yanson,Smit2003,Vardimon2014}. An oscillatory dependence of
the conductance with the length of gold atomic chains has been
reported by Smit \textit{et al.} \cite{Smit2003}, suggesting a
dependence on whether the number of atoms in the chain is even or
odd and originating from interference effects in the electron wave
functions. However, Vardimon \textit{et al.} \cite{Vardimon2014},
carried out conductance and shot noise measurements, and inferred,
consistent with theory \cite{Tavazza13}, that the conductance
oscillations of gold atomic chains during stretching are mainly
related to variations in bond angles (and hence in orbital
overlaps) as the chain undergoes transitions between zigzag and
linear atomic configurations.

The conductance properties of gold atomic chains have also been
investigated theoretically by means of extended H\"{u}ckel theory
\cite{Emberly}, classical molecular dynamics simulations combined
with a tight-binding approximation \cite{Dreher}, and \textit{ab
initio} calculations \cite{Vega,Tavazza,Tavazza13,Zheng2015}. It
was shown that the conductance of gold atomic wires is close to
1$g_0=2e^2/h$ for wires with different lengths
\cite{Emberly,Vega}. In addition, the gold chains are
characterized by a single conducting channel around the Fermi
energy with predominant half-filled $s$ character, giving rise to
the even-odd conductance oscillation with the number of atoms in
the chain due to quantum interference \cite{Vega} or conductance
oscillations due to bond angle and bond length changes during
elongation of zigzag atomic chains. \cite{Tavazza13} A similar
mechanism involving changing bond angles has been proposed by us
to account for variation of the conductances of molecular
junctions bridging gold electrodes in response to their elongation
under tensile stress \cite{Saffarzadeh14}. In the present work we
study the effects of both mechanisms, those of quantum
interference by varying the number of atoms in the chain while
keeping the average interatomic spacing fixed, and those due to
variation of bond angles and bond lengths by stretching the atomic
chains while keeping the number of atoms in them fixed. We find
both mechanisms to influence the conductance.

On the other hand, the thermoelectric properties of atomic and
molecular junctions are particularly interesting for developing
high efficiency energy conversion devices
\cite{Reddy,Giazotto,Zimbovskaya,Tsutsui1,Sadeghi,Perroni,Garcia,Mosso,Aiba1,Widawsky2013,Kim2014,Cui2017}.
Owing to the advances in experimental techniques developed to
study the thermoelectric properties of atomic and molecular
bridges \cite{Ludoph,Reddy,Garcia,Tsutsui2}, much progress has
been achieved in understanding and controlling electron and heat
transport properties of atomic contacts \cite{Mosso,Cui2017}.
Recently, it has been demonstrated that the thermopower of
metallic atomic-size contacts can differ qualitatively from that
in the macroscopic limit \cite{Evangeli}, such that few-atom gold
and platinum junctions exhibit opposite signs in their
thermopower, unlike bulk structures. This difference is due to the
discreteness of the energy levels responsible for transport in
atomic junctions (6$s$ orbitals for gold and 5$d$ orbitals for
platinum) \cite{Evangeli,Pauly} and the tunability of their
properties by external factors such as pressure and electrostatic
gates \cite{Garcia}. A sign change of the thermoelectric voltage
has also been reported in helicene molecular junctions with an
on/off ratio of more than three orders of magnitude \cite{Vacek}.

More recently, we have reported bipolar thermoelectric voltage
generation in few-atom gold contacts, achieved by controlling
electronic quantum interference through the application of
mechanical force \cite{Aiba1}. Despite its potential importance
for energy conversion nanoelectronics, the present-day
understanding of the relationship between the polarity change of
thermoelectric voltage and the junction length or the number of
atoms in atomic chains bridging two metal electrodes and also the
role of the atomic layer structures within the electrodes of
atomic-sized junctions require further investigation.
\begin{figure}
\centerline{\includegraphics[width=0.9\linewidth]{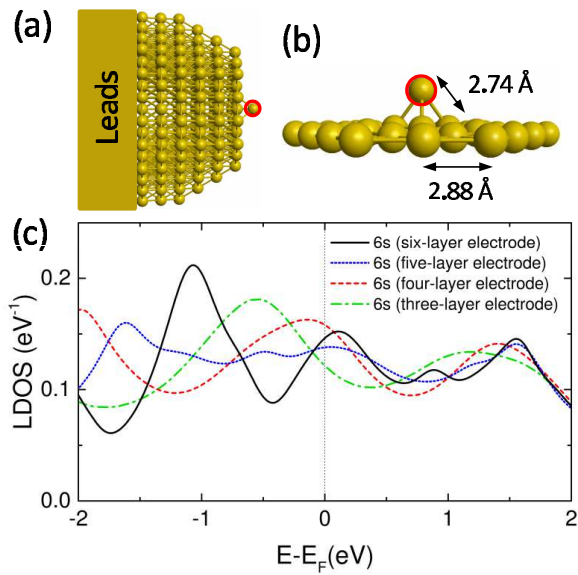}}
\caption{(Color online) (a) The side view of a six-layer electrode
in contact with semi-infinite leads. (b) Optimized geometry of the
tip atom and the first layer of the electrode. The red circle
denotes the tip atom. (c) Calculated local density of states of
tip $6s$ orbital in the vicinity of Fermi energy for three-,
four-, five-, and six-layer electrodes, obtained by extended
H\"{u}ckel semi-empirical calculations.} \label{F1}
\end{figure}

In order to shed light on this issue, in this paper the atomic
chain length-dependence of the electron transmission probability
and thermoelectric voltage of gold atomic wire junctions with
regular and irregular electrode grain boundaries are explored
theoretically by means of density functional theory (DFT)
calculations and a semi-empirical tight-binding method based on
the extended H\"{u}ckel theory \cite{Geo,Ammeter,YAEHMOP}. A
heuristic interpretation of the numerical results is also
provided. The junction length is varied either by adding atoms to
atomic chain bridging two electrodes or by imposing tensile
strains while keeping the number of atoms in the chain fixed. We
present our theoretical analysis of thermoelectric effects in
structures consisting of a few hundreds of atoms ranging from 225
(three-layer electrodes and a single Au atom in between) to 607
(six-layer electrodes bridged by a 21-atom chain). Our electrical
and thermal transport calculations show that the thermoelectric
voltages of these atomic wire junctions exhibit regular
oscillations (with successive sign changes and growing amplitude)
as the number of atoms in the atomic chain is increased.
Furthermore, we demonstrate that the polarity of the
thermoelectric voltage can be controlled mechanically by applying
a tensile strain.

The paper is organized as follows. In Sec. II, we outline the
computational details of the energy optimization of atomic
junctions and of the electrical and thermal transport
calculations. In Sec. III, we discuss the results for junctions
with differing numbers of atoms in the chains bridging the
electrodes, the influence of electrodes with differing numbers of
atomic layers, and also the effect of regular and irregular grain
boundaries. In addition, we present our results for the influence
of mechanical strain on the total energy, transmission
probability, and thermoelectric voltage for typical junctions.
Finally, in Sec. IV, we conclude this work with a general
discussion of the results.

\section{Theory}
\subsection{Modeling the structures}
In the models that we consider here, the electrodes are
represented by a pair of atomic clusters with up to several
hundred gold atoms and are bridged by atomic chains ranging from 1
to 21 gold atoms that form the junction. The positions of the gold
atoms in the atomic chain and in its immediate vicinity were
estimated by minimizing the total energy of the system, computed
within DFT using the GAUSSIAN 16 package with the PBE1PBE
functional and Lanl2DZ pseudopotentials and basis sets
\cite{gaussian,Pedrew}. The gold atoms of the electrodes that are
further from the junction are fixed during the relaxations and are
assumed to have the geometrical configuration of the ideal fcc
lattice in the crystal direction $<111>$ with nearest neighbor
distance 2.88 {\AA}.

For our studies of the dependence of the thermoelectric voltage on
the number $N$ of atoms in the atomic chain we constructed the
chains atom by atom, increasing the separation of the atomic
clusters representing the electrodes by 2.497\AA~each time that an
atom was added to the chain and then relaxing the structure as
described above. For our studies of dependence of the
thermoelectric voltage on the tensile strain in the junction, we
started with the structures obtained as above for the studies of
the $N$-dependence and then increased the separation between the
electrodes in small steps, relaxing the positions of the gold
atoms in the chain and in its immediate vicinity, while holding
the other atoms of the electrodes frozen.

Our DFT calculations directed at finding the energetically
favorable location of the tip atom on the surface of each gold
electrode reveal that the hollow site configuration is the most
stable, compared to the top and bridge sites. This result was
obtained not only for a single Au atom on the surface of gold
electrodes, but also for each gold atomic chain in contact with
electrodes. Further, the surface structure did not change
drastically during the optimization process, indicating that of
the electrode atoms only those in the vicinity of the tip atoms
need to be included in optimizing the geometry of the gold
nanojunctions. In order to determine how many neighbors of the tip
atom of each electrode should be considered unfrozen in the energy
minimization, we have optimized the first atomic layer of each
electrode plus tip atom, while only the positions of the
surrounding (outermost) atoms in the layer were fixed.

In Fig. 1(a) we show a side view of a single electrode consisting
of six atomic layers in which all orbitals belonging to the sixth
layer are connected to one dimensional leads. The optimized
structure of the tip atom (marked by a red ring) and the atomic
arrangement in the first layer are also shown in Fig. 1(b). The
bond length between tip and its immediate (electrode) neighbors
was found to be 2.74 {\AA} which is less than in plane Au-Au bond
length. It is clear that although all the interior atoms in the
layer are free to move, the positions of the three immediate
neighbors of the tip atom have changed slightly and it is
energetically favorable for these atoms to move out of the plane
towards the tip atom. For this layer, we found at most 0.022 {\AA}
atomic displacement in the plane and 0.34 {\AA} out of the plane.

\subsection{Thermoelectric calculations}
In order to investigate the effects of the gold clusters
(electrodes) and atomic chain sizes in the aforementioned
nanojunctions on the thermoelectric properties, we first calculate
the electron transmission probabilities $\mathcal{T}$ through the
device (atomic chain and electrodes) at the Fermi energy $E_F$. To
do this, we attached a large number of semi-infinite
quasi-one-dimensional ideal leads representing electron reservoirs
to the valence orbitals of the atoms in the outer layer of each of
the gold clusters that represent the two electrodes, as in
previous work on transport through nanostructures
\cite{Firuz1,Cardamone,Piva1,Piva2,Dalgleish1,George2,Saffarzadeh14}.
In the present model, the boundaries between these ideal leads and
the gold clusters may be viewed as electrode grain boundaries that
play a role in the thermoelectric behavior of experimental devices
\cite{Aiba1}. The role of such grain boundaries which can be
regular or irregular in our model will be discussed in the next
section. The electron transmission probability $\mathcal{T}(E)$
through the system, within Landauer theory \cite{Geo}, is related
to the conductance via $g(E_F)=g_{0}\mathcal{T}(E_F)$, and can be
written as
\begin{equation}\label{T}
\mathcal{T}(E_F)=\sum_{\alpha,\beta}|t_{\beta\alpha}(E_F)|^2\frac{v_\beta}{v_\alpha}\
.
\end{equation}
Here $t_{\beta\alpha}$ is the transmission amplitude through the
entire device and $\alpha$ ($\beta$) is the electronic state of a
carrier with velocity $v_\alpha$ ($v_\beta$) in the electron
source (drain) leads.

On the other hand, it is known that if a temperature difference
$\Delta T$ is imposed between two electrodes bridged by a
nanoscale junction, the thermoelectric voltage $\Delta V$ between
the electrodes (with no electric current flowing through the
junction) is given by\cite{PD}
\begin{equation}\label{Sc}
\Delta V = \frac{\pi^2 k^2_B T}{3e}\frac{1}{
\mathcal{T}(E)}\left.\frac{\partial \mathcal{T}(E)}{\partial
E}\right|_{E=E_F}\Delta T\  ,
\end{equation}
where $k_B$ and $e$ are the Boltzmann constant and the electron
charge, respectively, and $\mathcal{T}(E)$ is the electron
transmission probability at energy $E$ through the device that
includes the atomic wire and both electrodes. Note that Eqs.
(\ref{T}) and (\ref{Sc}) are valid provided that the electron
transmission probability varies smoothly in the range
$|E-E_F|<k_BT$ (as will be seen below in Fig. (2)) and that the
temperature difference $\Delta T$ between the electrodes is much
smaller than the mean temperature $T$. In addition, it should be
mentioned that the above equations do not include
electron-electron interactions beyond mean field theory, or
electron-phonon interactions, i.e. phonon drag. In macroscopic
metallic wires the phonon drag contribution to the thermopower is
important at low temperatures primarily due to the interaction of
electrons with long wavelength acoustic phonons \cite{Weber1996}.
However, for metallic atomic point contacts the phonon drag
contribution is believed to be suppressed \cite{Weber1996,Pauly}.
Also, because gold atomic chains are one-dimensional
single-channel electronic systems, absorption of long wavelength
acoustic phonons by electrons is expected to be suppressed in them
due to energy and crystal momentum selection rules. Consequently
the role of electron-phonon interactions will not be discussed
further in the present work. Here we focus on the electron
contribution to the coherent transport within the Landauer
formalism \cite{Garcia}.

When an electron traverses the device, it is scattered multiple
times from the surfaces of the electrodes as well as from grain
boundaries and also from the boundaries between the atomic wire
and electrodes. Consequently, the thermoelectric voltage $\Delta
V$ given by Eq. (\ref{Sc}) can be affected by all of these
structural details as well as the electron energy and $\Delta T$
\cite{Ludoph}. In this context it is worth mentioning that in
experiments, other defects such as dislocations and vacancies that
the electrodes may contain can also play a role in the
thermoelectric properties of devices.
\begin{figure}
\centerline{\includegraphics[width=1.0\linewidth]{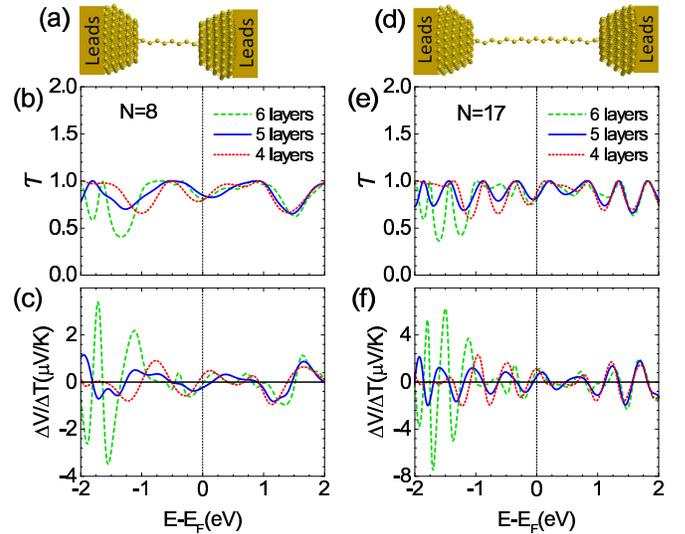}}
\caption{(Color online) (Optimized gold atomic wire junctions
consisting of (a) $N=8$ and (d) $N=17$ atoms with five-layer
electrodes in contact with semi-infinite leads. (b, e) Calculated
Landauer electron transmission probabilities and (c, f)
thermoelectric voltages versus electron energy $E$ around the
Fermi energy for four-, five- and six-layer electrodes at a mean
temperature of 27 K. Note that the thermopower or Seebeck
coefficient $S$ is related to the quantities shown here by
$S=-\Delta V/\Delta T$.} \label{F2}
\end{figure}

\section{Results and discussion}
The electronic states of the tip atoms influence electron
conduction through the gold nanojunctions. Transport experiments
on Au chains have shown that among the gold atomic valence
orbitals that include $5d$, $6s$, and $6p$ states, the electronic
density of states around the Fermi energy is dominated by $6s$
orbitals, indicating that the conduction channel is mainly a
single $6s$ band \cite{Rego,Agrait,Evangeli}. In Fig. 1(c) the
$6s$ orbital-tip density of states is depicted for single
electrodes consisting of different numbers of atomic layers. We
see that although the local density of states changes as the
number of layers in the gold electrode increases from three to
six, the change is not significant at Fermi energy.

\begin{figure}[ht]
\begin{center}
\includegraphics[scale = 0.36]{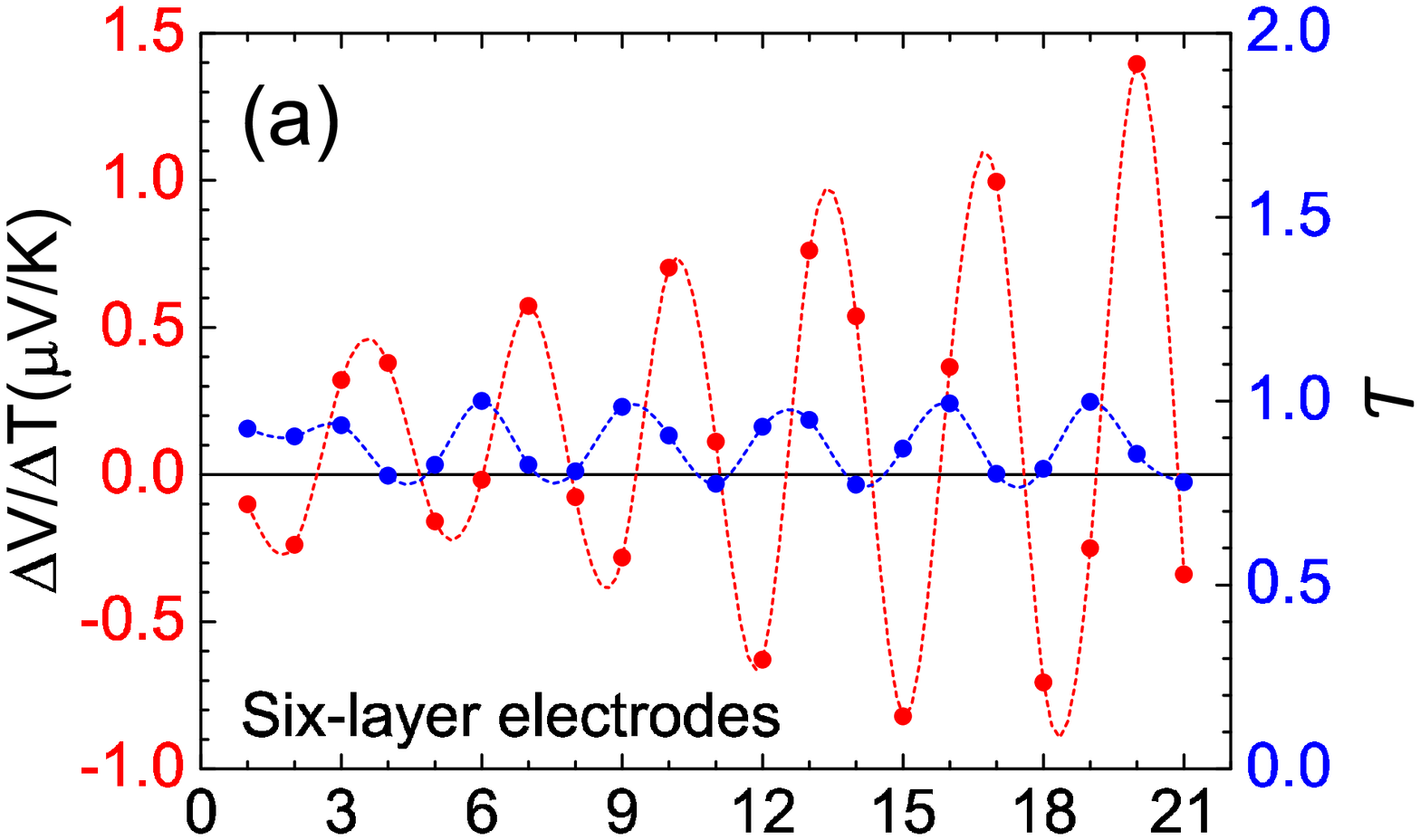}
\\
\vspace{0.03in}
\includegraphics[scale = 0.36]{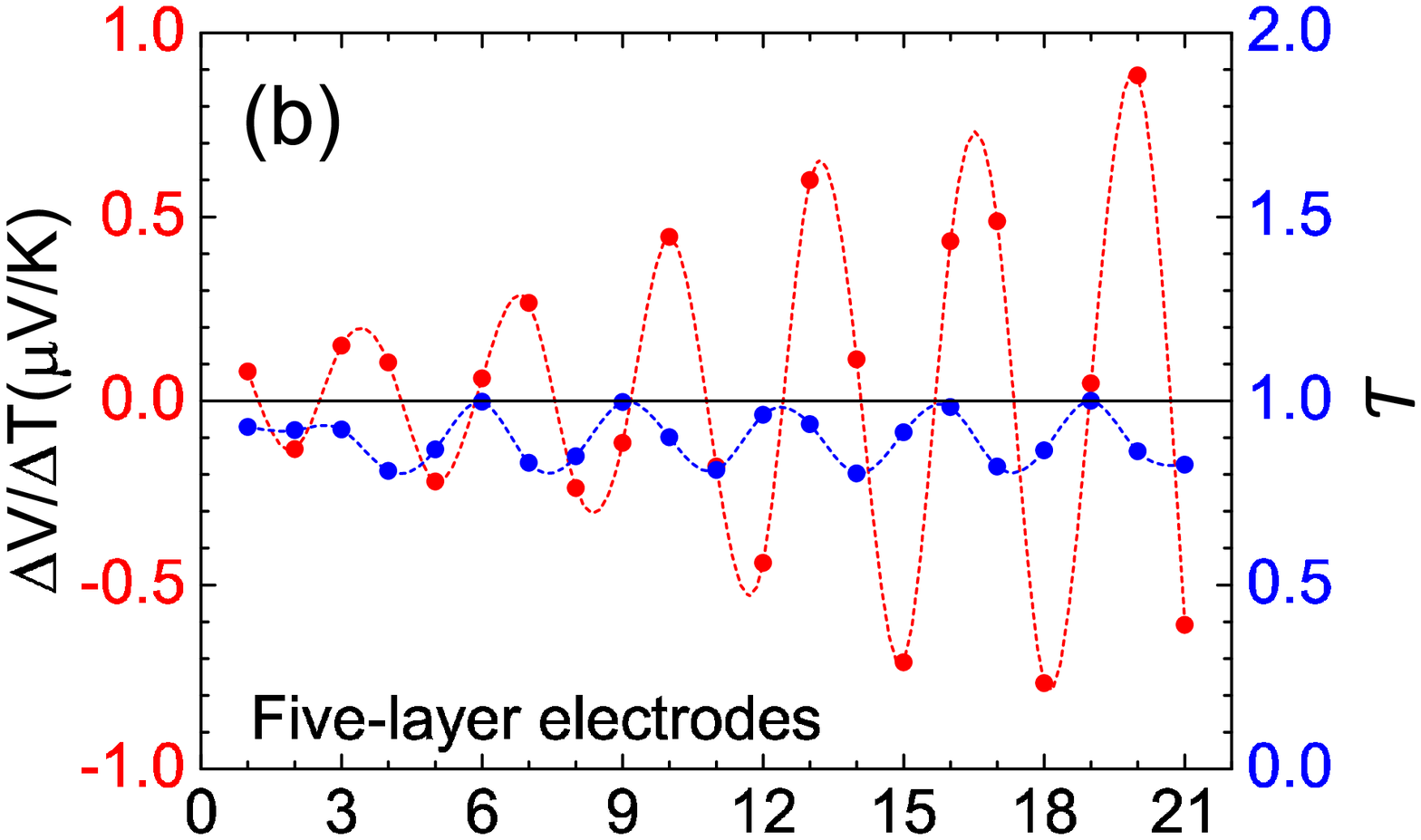}
\\
\vspace{0.03in}
\includegraphics[scale = 0.36]{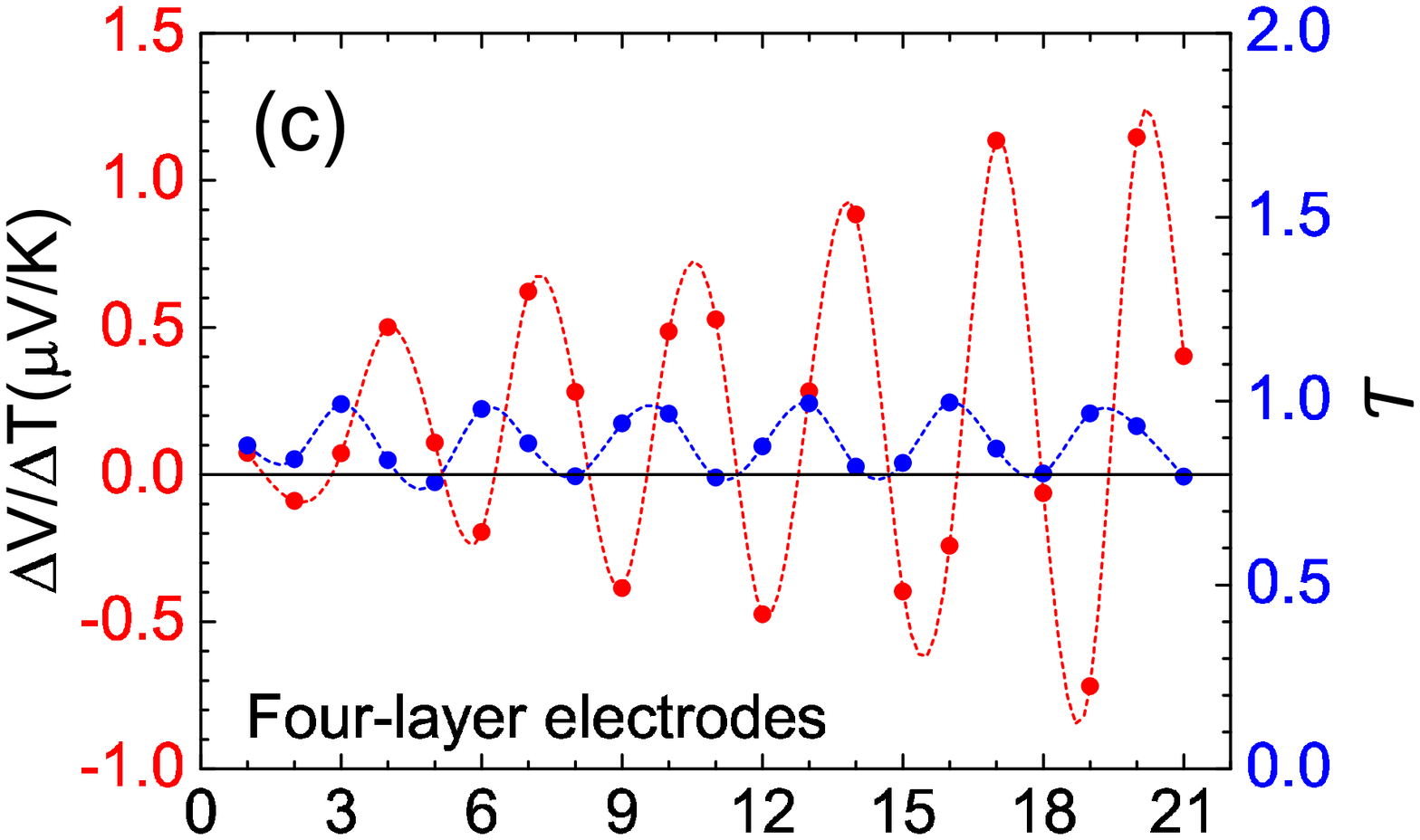}
\\
\vspace{0.03in}
\includegraphics[scale = 0.36]{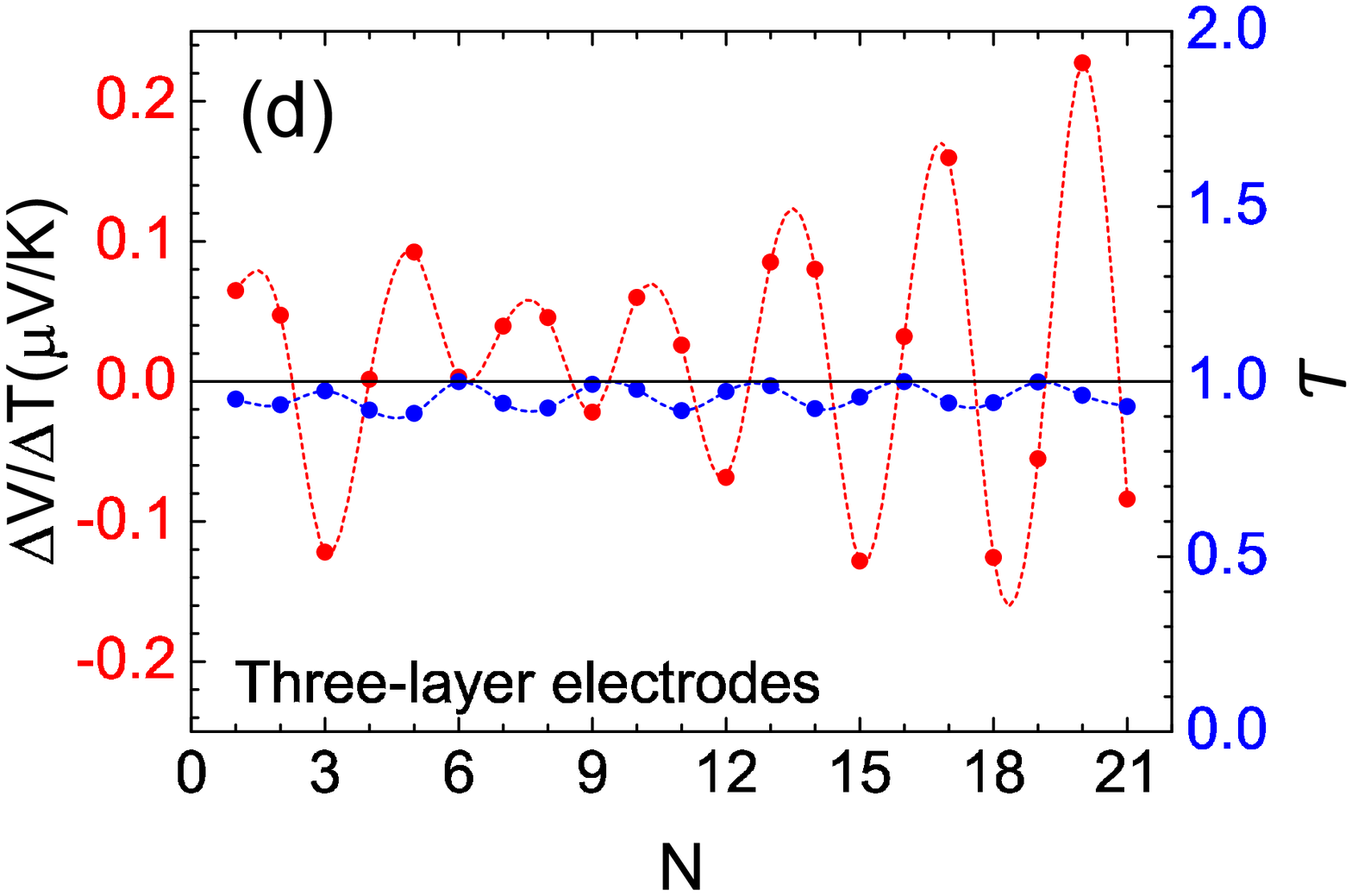}
\caption{(Color online) Calculated electron transmission
probabilities $\mathcal{T}$ (blue dots) and thermoelectric
voltages (red dots) at the Fermi energy versus the number of gold
atoms $N$ in the wire junctions bridging the two electrodes, each
consisting of (a) six, (b) five, (c) four, and (d) three gold
atomic layers at a mean temperature of 27 K. Dotted lines are
guides to the eye. }\label{F3}
\end{center}
\end{figure}

We will now consider atomic chains with different numbers $N$ of
gold atoms sandwiched between two gold electrodes, each of which
may contain different numbers of atomic layers. All the atomic
wire junctions (i.e., the two electrodes and the atomic wire in
between) are optimized in the same way as described in Sec. IIA.
In Fig. 2, we have depicted the transmission probability and
thermoelectric voltage versus electron energy, $E$, for the wire
junctions with $N=8$ and $N=17$ and four- to six-layer electrodes.
The peaks in the electron transmission probability $\mathcal{T}$
are due to organ pipe-like electron standing waves forming in the
device and giving rise to transmission resonances
\cite{Emberly,Aiba1,Kirczenow1989}. We see that the number of
oscillations in $\mathcal{T}$ and $\Delta V/\Delta T$ increases as
the number of atoms in the atomic chain is increased. Indeed, the
number of transmission resonances at which $\mathcal{T}(E)=1$ (in
the energy range shown) increases when more atoms are added to the
chain.

The underlying reason for this is that for a longer atomic chain a
smaller change in the electron de Broglie wavelength
(corresponding to a smaller change in the electron energy $E$) is
sufficient to switch the device between adjacent electronic
standing wave resonances, i.e., between adjacent peaks of
$\mathcal{T}(E)$ in Fig. 2(b) and (e). This can be understood
intuitively by analogy with the acoustic resonances of a pipe
which obey the resonance condition $L = n \lambda/2$ where $L$ is
the effective length of the resonator, $\lambda$ is the resonant
wavelength and $n$ is an integer. The difference between the
wavelengths $\lambda$ and $\lambda'$ of two consecutive resonances
can then be written as $\Delta\lambda=\lambda-\lambda' =
\frac{\lambda\lambda'}{2L}$. Thus for large $L$ the difference
between the energies of two consecutive resonances is $\Delta E
\approx \frac{\partial E}{\partial \lambda}\Delta\lambda \approx
\frac{\partial E}{\partial \lambda}
\frac{\lambda_\text{dB}^2}{2L}$ where $\lambda_\text{dB}$ is the
electron de Broglie wavelength at the Fermi level. Thus for large
numbers $N$ of atoms in the atomic chain $\Delta E \propto 1/N$.

This behavior of the electron transmission probability
$\mathcal{T}(E)$ in turn affects the quantity $\Delta V/\Delta T$.
Importantly, the fact that oscillations in $\mathcal{T}(E)$ become
more closely spaced in energy $E$ with increasing number $N$ of
atoms in the chain results in an increase in $\frac{\partial
\mathcal{T}(E)}{\partial E}$ with increasing $N$ that is reflected
in an increase in the thermoelectric voltage $\Delta V$ given by
Eq. (2). In particular, if  $\mathcal{T}(E)$ can be regarded as
sinusoidal and the amplitude of the oscillations in
$\mathcal{T}(E)$ is not sensitive to $N$ [as can be seen by
comparing Fig. 2(b) and (e)], then $\frac{\partial
\mathcal{T}(E)}{\partial E}$ is inversely proportional to the
energy spacing  between the peaks of $\mathcal{T}(E)$. Then our
result $\Delta E \propto 1/N$ obtained above translates into the
amplitude of the oscillations of $\Delta V/\Delta T$ being
directly proportional to the number $N$ of gold atoms in the chain
for large $N$. This heuristic prediction is consistent with the
increase in the amplitude of the oscillations of $\Delta V/\Delta
T$ by approximately a factor of 2 from Fig. 2(c) [$N$=8] to Fig.
2(f) [$N$=17]; note the factor 2 difference between the vertical
scales of the two figures. It is further supported strongly by the
systematic studies of the dependence of the thermoelectric voltage
on the number of atoms $N$ in the atomic chains that will be
presented below.

In the case of $N=8$, as the number of atomic layers in the
electrodes increases from 4 to 6, $\Delta V/\Delta T$ changes from
positive to almost zero to negative values at Fermi energy (see
Fig. 2(c)). This behavior is totally different than that for the
case of $N=17$ where $\Delta V/\Delta T$ is positive at the Fermi
energy for all three sizes of the electrodes (see Fig. 2(f)).
These results suggest that the sign of the thermoelectric voltage
can be affected by the number of layers in electrodes as well as
by the number of atoms in the chain.

To better understand the role of $N$ and the number of atomic
layers in the electrodes, we have shown in Fig. 3 the quantities
$\mathcal{T}$ and $\Delta V/\Delta T$ at the Fermi energy as a
function of the number of chain atoms $N$ in wire junctions with
different electrodes. It is evident that both quantities oscillate
as $N$ changes. However, while the amplitude of the oscillations
of $\mathcal{T}$ is insensitive to $N$, the amplitude of the
oscillations of $\Delta V/\Delta T$ increases proportionally to
$N$ for larger values of $N$, as was predicted heuristically
above. Although for small and moderate $N$ the amplitude of
oscillations of the thermoelectric voltage are irregular in the
structures with three-layer electrodes (see Fig. 3(d)), they
become rather regular if more atomic layers are included in the
electrode structures (see Figs. 3(a)-(c)). Nevertheless, small
deviations from regular oscillations in transmission probabilities
occur for the short junction lengths with $N=1-3$. Indeed, the
nature of atomic chains in such nanojunctions manifests itself for
$N\geq 4$. This result makes sense because in all of the wire
junctions, the first and the last atoms of the chain are bonded to
the electrodes through three atomic bonds, while the other chain
atoms make only a single bond with their neighbor atoms.
Importantly, the sign of the thermoelectric voltage oscillates
between positive and negative values with $N$, regardless of the
number of atomic layers in the electrodes. The sign changes in the
thermoelectric voltage with increasing numbers of layers or chain
atoms reveal an important property of gold atomic junctions and
suggest a potential application as a perfect voltage switch. In
other words, the polarity of the voltage can be switched by
elongation of the junction. In the junction with five-layer
electrodes, the oscillations of $\Delta V/\Delta T$ between
positive and negative values versus $N$ are almost symmetric,
while there is a shift towards more positive values for the
junctions with four- and also six-layer electrodes (see Figs. 3(a)
and (c)), which can be attributed to the quantum interference of
electrons in the electrodes.

Sign differences of the thermoelectric voltage have been observed
previously in atomic-scale and molecular junctions. For instance,
amine-terminated molecular junctions have positive thermoelectric
voltages, whereas pyridine-terminated molecular junctions exhibit
negative values of their thermoelectric voltages \cite{Widawsky}.
However, the oscillatory characteristics of $\Delta V/\Delta T$
with increasing amplitudes and the polarity changes for long
atomic chain junctions predicted here have not been reported
previously. Also, it has been observed that gold atomic-sized
contacts have an average negative thermopower, whereas platinum
contacts present a positive thermopower \cite{Evangeli}. By
contrast, we have shown that both positive and negative polarities
of the thermoelectric voltage can be generated in an atomic wire
junction. This feature improves the feasibility of realizing
voltage switches based on nano-wire junctions.

\begin{figure}
\centerline{\includegraphics[width=0.85\linewidth]{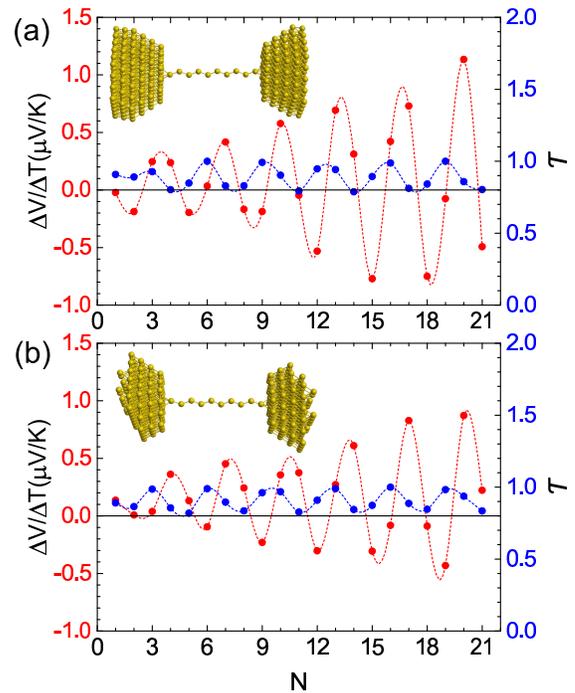}}
\caption{(Color online) Calculated electron transmission
probabilities $\mathcal{T}$ (blue dots) and thermoelectric
voltages (red dots) at the Fermi energy versus the number of gold
atoms $N$ in the wire junctions between (a) a six-layer electrode
at left and a five-layer electrode at right with regular grain
boundaries and (b) five-layer electrodes with irregular grain
boundaries at a mean temperature of 27 K. The insets show two
typical gold atomic junctions with $N=10$. The semi-infinite leads
are not shown here. Dotted lines are guides to the eye.}
\label{F4}
\end{figure}

In the structures that we have considered so far, the interfaces
between the ideal leads and electrodes that can be regarded as
representing crystal grain boundaries were regular and the two
electrodes had the same number of atomic layers. In Fig. 4 (a),
however, we have depicted the electron transmission probability
and thermoelectric voltage for the wire junctions with different
electrodes, i.e., a six-layer electrode at left and a five-layer
electrode at right. We see that the oscillations in both
$\mathcal{T}$ and $\Delta V/\Delta T$ are still regular and the
amplitude of thermoelectric oscillations increases nearly linearly
with $N$ as the number of atoms $N$ in the chain is increased.
Moreover, the oscillations of $\Delta V/\Delta T$ are almost
symmetric with respect to the zero line. This suggests that
switching to gold atomic wire junctions with electrodes having
differing numbers of atomic layers and regular grain boundaries
does not affect the sign change mechanism of thermoelectric
voltage. In the case of irregular grain boundaries, however, the
amplitude of the oscillations of $\Delta V/\Delta T$ decreases
somewhat and the thermoelectric voltage becomes skewed to positive
values as shown in Fig. 4(b). Nevertheless, the regularity of the
oscillations, the growth of their amplitude with increasing $N$
and the occurrence of sign changes of $\Delta V/\Delta T$ are, for
the most part, not affected by the irregular grain boundaries.

Cortes-Huerto \textit{et al.} \cite{Cortes-Huerto} have shown that
the growth of monatomic chains can stop if the contact tips become
symmetric because the system requires more energy to remove one
gold atom from the contacts and add it to the chain than that
needed for bond breaking within the chain. We note, however, that
although in this study the atomic junctions extend along $<111>$
crystal direction, the gold clusters do not have the same symmetry
and thus, it is expected that long atomic chains may form before
the junctions rupture. Although long atomic chains with up to 21
atoms may not have been realized experimentally at this time, our
\textit{ab initio} calculations and semi-empirical extended
H\"{u}ckel model suggest that such long chains can be formed
between two atomic-size contacts and should have interesting
properties. This may motivate experimentalists to make such long
atomic chains and measure their transport characteristics.

\begin{figure}
\centerline{\includegraphics[width=0.95\linewidth]{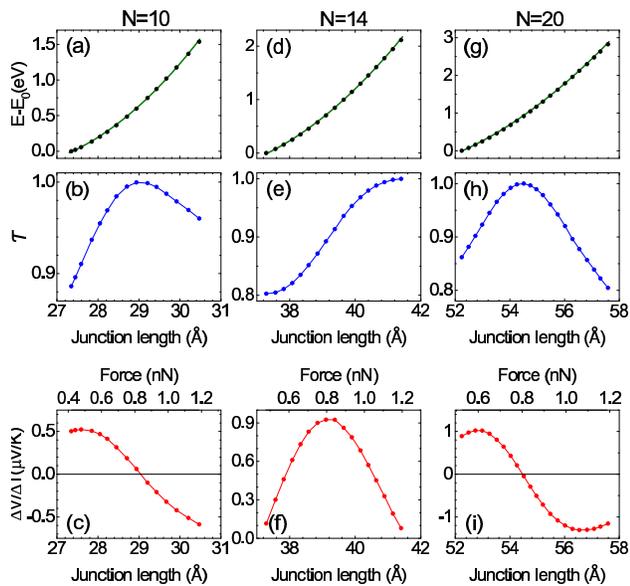}}
\caption{(Color online) Calculated [(a), (d), and (g)] total
energies, [(b), (e), and (h)] electron transmission probabilities,
and [(c), (f), and (i)] thermoelectric voltages at the Fermi
energy versus the junction length in the wire junctions with
five-layer electrodes and regular grain boundaries at a mean
temperature of 27 K. The top axis in (c), (f), and (i) shows the
force required for elongation of the junction. The green curves
show the fitted results, while the blue and red lines are guides
to the eye. $E_0$ represents the total energy at zero strain.}
\label{F5}
\end{figure}

The response of thermoelectric properties to variable tensile
strains, however, is of great importance for both scientific
understanding and potential applications such as atomic-sized
switches and high-performance energy conversion devices
\cite{Smith,Rubio1996,Mosso}. In this context, we have also
investigated the effect of junction elongation in the structures
with atomic chains having fixed numbers of atoms, specifically,
chains with $N=10$, $N=14$, and $N=20$. As shown in Figs. 5(a),
(d), and (g), by increasing the tensile strain, the total energy
of all structures increases quadratically, until the chains
rupture. On the other hand, the magnitude of mechanical force
required to elongate the junctions can be obtained as the
derivative of the total energy with respect to the junction
length. For this purpose, precise quadratic fits to the total
energy diagrams in the range of elongation lengths are employed
(see the green curves in Figs. 5(a), (d), and (g)).

Although the transmission probability increases initially with
junction elongation in all structures (see Figs. 5(b), (e), and
(h)), it may reach its maximum value and decrease again (see Fig.
5(b) and (h)). This behavior manifests itself as a sign change of
the thermoelectric voltage as shown in Figs. 5(c) and (i). In
other words, when the transmission probability passes its maximum
value, the sign of thermoelectric voltage switches from positive
to negative (see Figs. 5(c) and (i)). However, such a maximum does
not exist for the structure with $N=14$. Indeed, in Fig. 5(e), the
transmission probability increases with junction length until the
chain ruptures. We see that if the chains bridging the electrodes
support the external strains, both $\mathcal{T}$ and $\Delta
V/\Delta T$ can show partial oscillations. In Figs. 5(c), (f), and
(i) we have also shown the range of forces required for elongation
of the junctions. Interestingly, the maximum mechanical force does
not exceed 1.2 nN, irrespective of the chain length. In this
respect, Rubio \textit{et al.} \cite{Rubio2001} showed
experimentally that the force required to break one single bond in
the chain is 1.5 nN which is larger than the maximum value 1.2 nN
in this study. This suggests that junction elongations similar to
those shown in Fig. 5 can be achieved experimentally and thus the
sign changes of the thermoelectric voltage predicted in Fig. 5(c)
and (i) should be observable.

In order to demonstrate that our theory is able to reproduce the
results of experimental conductance measurements, we compare the
oscillations of the transmission probabilities versus $N$ in Figs.
3 and 4 with the evolution of the conductance versus chain length
in Refs. [\onlinecite{Smit2003}] and [\onlinecite{Vardimon2014}].
According to the experimental results, shown for Au in Fig. 2 of
Ref. [\onlinecite{Smit2003}] and also in Fig. 1(d) of Ref.
[\onlinecite{Vardimon2014}], the conductance of the last plateau
with values $\sim 1g_0$ shows oscillations in the process of
elongating the gold atomic chain and the electric conductance
changes by about 10-15\%. This result is in good agreement with
our theory in which the transmission probability ($g/g_0$) of the
gold wire junctions versus $N$ shows oscillations and its value
varies by about 10-20\% (see $\mathcal{T}$ versus $N$ in Figs. 3
and 4). We also find modulation of $g/g_0$ by 10-20\% when the
junction is stretched without changing $N$ (as in Fig. 5), again
in agreement with the experiments.

Previously reported calculations of Vega \textit{et al.}
\cite{Vega} for finite Au chains, obtained an oscillatory behavior
of the conductance due to a single conducting channel around the
Fermi energy with predominant $s$ character, which is in good
agreement with our results for $\mathcal{T}$ versus $N$ in Figs. 3
and 4. Furthermore, as we mentioned above, the variation of
conductance with chain length in our theory is in good agreement
with experiments [\onlinecite{Smit2003}] and
[\onlinecite{Vardimon2014}], whereas the conductance oscillations
in the theory of Vega \textit{et al.} \cite{Vega} exhibit smaller
amplitudes $\sim 0.04g_0$, suggesting that our modeling may be
more realistic. However, the amplitudes of the conductance
oscillations in the simulations of Tavazza \textit{et al.}
\cite{Tavazza13} were similar to those found in the present work.

Using an analytic approach to study transport in a one-dimensional
(1D) atomic chain with one orbital per site as a model of a single
channel wire we have also shown that the transmission probability
and thermoelectric voltage (see appendix A) exhibit regular
oscillations due to the quantum interference effects in the
electron wave functions originating from electron scattering from
the barriers between the $N$-atom central chain and the
electrodes. Comparing the aforementioned results obtained by
combining DFT and extended H\"{u}ckel parameters with those given
in Fig. 6, we conclude that the details of the bonding between the
electrodes and the atomic chain, of the atomic arrangements in the
electrodes, and of the multiple valence orbitals of each gold atom
which are all absent in the simplified 1D model studied
analytically in appendix A, result in significant deviations of
the oscillations from simple even-old oscillatory behavior with
the length of atomic chains.

Vardimon \textit{et al.} \cite{Vardimon2014} argued that the
conductance oscillations of gold atomic chains that they observed
experimentally were due to increasing conductance as zigzag chain
structures were straightened by stretching followed by decreasing
conductances when an additional gold atom was pulled out of a
contact and added to the chain. Tavazza \textit{et al.}
\cite{Tavazza13} found similar behavior theoretically and proposed
that a gradual decrease of the conductance can occur due to
increasing interatomic bond lengths. Our results for the
conductance in Fig. \ref{F5}(b), (e) and (h) where the atomic
chains are stretched while holding the number of atoms in the
chains fixed, agree well with these previous experimental and
theoretical findings. However, in order to observe  the regular
oscillations of  the conductance and thermoelectric voltage due to
quantum interference effects that we predict and are shown  in
Figs 3 and 4 it is necessary to separate them experimentally from
the effects due to formation and straightening of zigzag chain
structures and variation of interatomic bond lengths that mask
them in the conductance data of Vardimon \textit{et al.}
\cite{Vardimon2014}. This might be achieved by measuring the
conductances and thermoelectric voltages {\em only} of junctions
with atomic chains having equal average interatomic spacings but
different numbers of gold atoms, as in our calculations whose
results are reported in Figs 3 and 4. However, an alternative
approach that may be less challenging experimentally is to measure
the conductances and thermoelectric voltages for  atomic chains
having different numbers of gold atoms $N$ but subjected to the
{\em same} value of the applied tensile force. Thus all of the
studied atomic chains would have similar bond angles and bond
lengths, and any observed dependence of the conductance and
thermoelectric voltage on $N$ would be due primarily to quantum
interference, as in Figs 3 and 4.

Finally, we note that although some details of the results that we
have presented for the thermoelectric voltage depend on the exact
size and geometry of the contacts in our simulations, the
important aspects of our results are reasonably robust,
considering that we are dealing with nanoscale systems and
phenomena affected by quantum interference. The robust features
that we found are as follows: In all cases, for electrodes with
more than three atomic layers, we found regular oscillations of
the thermoelectric voltage with the amplitude of the oscillations
growing linearly with the number of atoms in the chain. Also, for
electrodes with more than three atomic layers, the period of the
oscillations was almost the same in all cases and the amplitude of
the oscillations (for equal numbers of atoms in the chains) varied
from system to system by less than a factor of two. This
robustness suggests that experiments studying thermoelectricity of
gold atomic chains may succeed in finding interesting results
despite the presence of defects such as disordered grain
boundaries in the contacts close to the atomic chain. However,
further theoretical studies with statistical analysis of the
results for large numbers of different sample geometries would be
of interest.

\section{conclusion}

In summary, using a combination of \textit{ab initio} and
semi-empirical calculations, we have presented a systematic
exploration of the response of the thermoelectric voltage of gold
atomic wire junctions to changes of the junction's length. The
junction's length was varied in two ways: (i) By successively
adding single gold atoms to the atomic chain bridging two gold
clusters, and (ii) by stretching the chain mechanically by
applying a tensile stress which increases the Au-Au bond lengths
and Au-Au-Au bond angles between atoms of the chain while keeping
the number of atoms in the chain fixed. Our results predict that
both the electron transmission probability through the junction
and the thermoelectric voltage should exhibit an oscillatory
behavior in response to adding more atoms to the atomic chain
bridging the electrodes. We predict the amplitude of the
oscillations of the thermoelectric voltage to increase in
proportion to the number of atoms in the atomic chain for long
atomic chains, based on our numerical results and on heuristic
reasoning. Similar behavior is also predicted if a variable
tensile strain is imposed provided that the atomic chain junction
supports the applied external force.

Our findings show that whether the junction elongation is imposed
by adding extra gold atoms to the atomic junction or by external
strain, a polarity change in thermoelectric voltage can be
achieved in the atomic metallic chains. This feature may enable
direct conversion of thermal energy into electricity and designing
mechanically controllable voltage switches \cite{Aiba1}.

\section*{Acknowledgement}
This work was supported by NSERC, CIFAR, WestGrid, and Compute
Canada.

\begin{appendix}
\section{Analytical calculation for transport through a single-channel wire}\label{Ap1}

The transmission function $\mathcal{T}(E)$ for a single-channel
wire can be calculated analytically by means of the
transfer-matrix formalism. We consider an infinite linear chain
with lattice constant $a$ described by a nearest neighbor
single-orbital tight-binding model with on-site energy
$\epsilon_0$ and the hopping integrals $\gamma$. To show the
evolution of $\mathcal{T}(E)$ versus the number of chain atoms
$N$, we assume the hopping parameter to be $0.8\gamma$ between
sites $l=0$ and $l=1$ and also between $l=N$ and $l=N+1$. This
means that the incoming electrons from left region (electrode) can
be partially transmitted to the right $(l\geq N+1)$ and partially
reflected back into the left region $(l\leq 0)$ due to the
electron scattering from sites $1\leq l\leq N$ acting as a
junction. The electron wave function can be written as
$|\Psi\rangle=\sum_l\psi_l|l\rangle$ where $|l\rangle$ represents
the single atomic orbital at site $l$. The Schr\"{o}dinger
equation for the coefficient $\psi_l$ at site $l$ is given by
\begin{equation}\label{Ap1}
(\epsilon_0-E)\psi_l+\gamma_{l,l+1}\psi_{l+1}+\gamma_{l,l-1}\psi_{l-1}=0\
,
\end{equation}
where $E$ is the electron energy. Within the transfer matrix
framework, Eq. (\ref{Ap1}) can be expressed as
\begin{eqnarray}\label{h0}
\left(%
\begin{array}{c}
\psi_{l+1} \\
\psi_{l} \\
\end{array}%
\right)
&=&\left(%
\begin{array}{cc}
\frac{E-\epsilon_{0}}{\gamma_{l,l+1}} &
-\frac{\gamma_{l,l-1}}{\gamma_{l,l+1}}\\
1 & 0 \\\end{array}%
\right)
\left(%
\begin{array}{c}
  \psi_{l} \\
  \psi_{l-1} \\
\end{array}%
\right)\nonumber\\
&=&T_l(E)
\left(%
\begin{array}{c}
  \psi_{l} \\
  \psi_{l-1} \\
\end{array}%
\right)\  .
\end{eqnarray}

Here, $T_l(E)$ connects the coefficients of the wavefunction at
site $l+1$ and $l$ to those at $l$ and $l-1$. Therefore, the
transfer matrix $P(E)$ connecting the sites $l=-1, 0$ to the sites
$l=N+1, N+2$ can be obtained from
\begin{eqnarray}\label{h1}
\left(%
\begin{array}{c}
\psi_{N+2} \\
\psi_{N+1} \\
\end{array}%
\right)&=&\prod^{0}_{l=N+1}T_l(E)
\left(%
\begin{array}{c}
  \psi_{0} \\
  \psi_{-1} \\
\end{array}%
\right)\nonumber \\
&=&P(E)
\left(%
\begin{array}{c}
  \psi_{0} \\
  \psi_{-1} \\
\end{array}%
\right)\  .
\end{eqnarray}
\begin{figure}

\centerline{\includegraphics[width=0.85\linewidth]{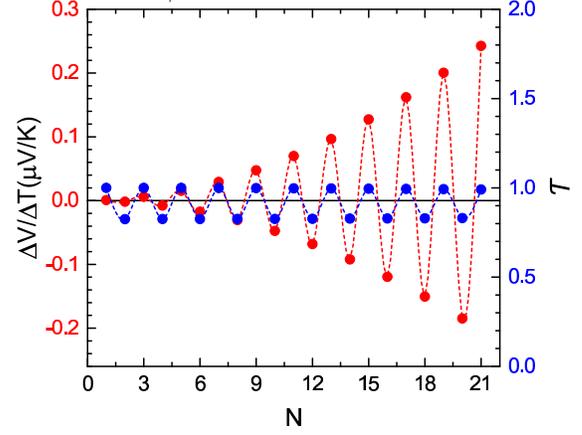}}
\caption{(Color online) Calculated electron transmission
probabilities $\mathcal{T}$ (blue dots) and thermoelectric
voltages (red dots) at $E=0$ versus the number of gold atoms $N$
in a single channel wire at a mean temperature of 27 K. Dotted
lines are guides to the eye.} \label{F6}
\end{figure}

The coefficients of wavefunction for $l\geq N+1$ and $l\leq0$ are
plane waves with wave vector
$ka=\cos^{-1}(\frac{E-\epsilon_0}{2\gamma})$. Therefore, the
incoming and outgoing scattering waves within the electrodes are
given by $\psi_{N+2}=\tau e^{ika(N+2)}$, $\psi_{N+1}=\tau
e^{ika(N+1)}$, $\psi_{0}=1+r$, and $\psi_{-1}=e^{-ika}+re^{ika}$
where $\tau$ and $r$ are the amplitudes of transmitted and
reflected wavefunctions. Using Eq. (\ref{h1}), the analytical form
of transmission probabilities $\mathcal{T}(E)=|\tau|^2$ for the
single-channel wire can be given by \small
\begin{equation}\label{tt}
\mathcal{T}(E)=\frac{4\sin^2(ka)}{[P_{12}-P_{21}+(P_{11}-P_{22})\cos(ka)]^2+(P_{11}+P_{22})^2\sin^2(ka)}\
.
\end{equation}
\normalsize

Therefore, the energy derivative of transmission probability,
$\partial\mathcal{T}/\partial E$, and hence the thermoelectric
voltage can be easily computed for this single-channel wire. The
results for $\mathcal{T}$ and $\Delta V/\Delta T$ are shown in
Fig. 6 with the parameters $\epsilon_0=0.05$ eV and $\gamma=-2.5$
eV.

As can be seen in Fig. 6, this simple analytic model reproduces
qualitatively the behavior obtained from the more detailed
numerical DFT and extended H\"{u}ckel-based approach presented in
the main text of this article. Namely, constant amplitude
conductance oscillations that are periodic in the chain length
$N$, and regular oscillations of the thermoelectric voltage whose
amplitude increases with $N$.

\end{appendix}

\end{document}